\documentclass[%
aip,
amsmath,amssymb,
reprint,%
]{revtex4-2}
\usepackage{graphicx}
\usepackage{dcolumn}
\usepackage{mathrsfs}
\usepackage{bm}
\usepackage{lineno}
\usepackage{hyperref}
\usepackage{bbm}
\usepackage{ulem}
\usepackage{color}

\begin{document}

\title{General considerations in unbalanced fourth-order interference}
\author{Z. Y. Ou}
 \email{jeffou@cityu.edu.hk}
\affiliation{Department of Physics, City University of Hong Kong, 83  Tat Chee Avenue, Kowloon, Hong Kong}

\author{Xiaoying Li}
 \email{xiaoyingli@tju.edu.cn}
\affiliation{College of Precision Instrument and Opto-Electronics Engineering, Key Laboratory of Opto-Electronics Information Technology, Ministry of Education, Tianjin University, Tianjin 300072, P. R. China}%

\begin{abstract}
Interferometry has been used widely in sensing application. However, the technique is limited by the finite coherence time of the light sources when the interference paths are not balanced. Higher-order interference effects involve intensity correlations between multiple detectors and may have the advantage over the traditional second order interference effect exhibited in only one detector.
We discuss various scenarios in fourth-order interference with unbalanced delays in different paths. We find in some cases, interference effect persists even when the delays are much larger than the coherence time of the sources. We also extend the discussion to non-stationary pulsed fields, which needs to consider the pulse shape and requires a different treatment. These results will be useful in remote sensing applications.

\end{abstract}

\maketitle

\section{Introduction}

Interferometry is the major technique for optical sensing applications \cite{opsens}. It depends on the optical coherence of light to produce phase-sensitive interference effect in order to achieve high sensitivity and precision \cite{bw}. This requires the balance of the interferometer paths to within the coherence length of the optical field. But this may limit the scope of applications in remote sensing when large imbalance of paths exists.

It was well-known that higher-order interference such as Hong-Ou-Mandel (HOM) interference effect \cite{hom,ou89} does not rely on the coherence time of the fields in that interference even between independent fields may occur \cite{xyli-th,Ma-th}. But such effect is insensitive to phase change of the fields. On the other hand, phase-dependent fourth-order interference effect occurs in Franson interferometer \cite{fran} which consists of two highly imbalanced interferometers beyond coherence length \cite{ou90,kwi,fran-ex91}. But it was shown that the effect exists only for two-photon quantum fields and disappears for stationary classical fields \cite{ou-mandel90}. The progress on the interference with imbalanced paths was halted until recently when it was reported that phase dependent fourth-order interference between two thermal fields can appear in the time-resolved coincidence between two detectors even when the path imbalance of the interferometer is well beyond the coherence length of the fields \cite{njp16,ihn17}. This leads to a huge advantage over the traditional interferometers based on second-order interference where interference appears in one detector and requires path imbalance between interfering fields be smaller than the coherence length of the fields.

Furthermore, even though HOM interference effect is independent of phase difference, it relies on mode match between the two input fields for photon indistinguishability required by quantum interference. Thus the size of the effect is sensitive to the distortion of the wave forms of the input fields. This is especially the case when the fields are in the form of ultra-short pulses and  can be a tool for sensing the change of the optical paths in the medium of propagation \cite{zhouqiang,Ma-th}.

In this paper, we discuss various scenarios in four-order interference with different correlation between interfering fields and unbalanced delays in different paths. We find in some cases, interference effect persists even when the delays are much larger than the coherence time of the sources. We also extend the discussion to non-stationary pulsed fields, which require the overlap of interfering pulses. The paper is organized as follows. We start discuss in Sec.II the general schemes with stationary fields. In Sec.III, we consider some special scenarios with different correlation between interfering fields and different delays for unbalanced interferometers. We consider the pulsed non-stationary fields in Sec.IV and conclude with a discussion in Sec.V.

\section{The case of stationary fields}

\begin{figure}[t]
\includegraphics[width=8cm]{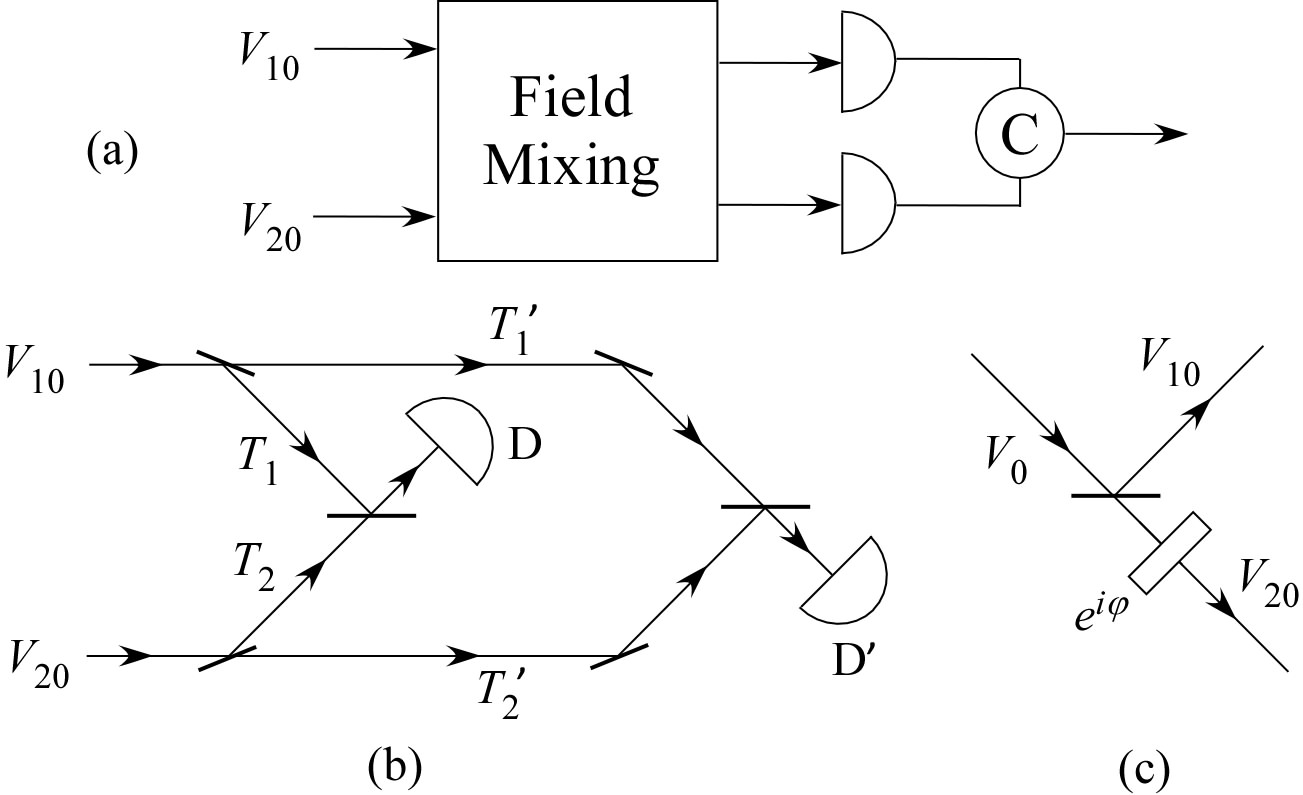}
	\caption{(a) General scheme of fourth-order interference between two fields by field mixing. (b) A specific scheme of field mixing by using beam splitters. (c) Generation of two fields with a random phase $\varphi$ from one field by a beam splitter.}
	\label{fig:sch}
\end{figure}

We start by considering fourth-order interference between two stationary fields $V_{10}({\bf r},t), V_{20}({\bf r},t)$. The quantities involved are related to the product of four field amplitudes such as $\langle V_{10}^*({\bf r}_1,t_1)V_{20}^*({\bf r}_2,t_2)$ $V_{10}({\bf r}_1',t_1')V_{20}({\bf r}_2',t_2')\rangle$. This requires intensity correlation between two different detectors. To achieve this,
we first mix the fields with some linear optics and send the mixed fields to two detectors for intensity correlation measurement, as shown in Fig.\ref{fig:sch}(a). The simplest way of field mixing is by beam splitters. A typical scheme is shown in Fig.\ref{fig:sch}(b), which will be the scheme of our discussion in this paper. In order to concentrate on fourth-order effect and avoid the confusion with lower order interference, that is, the interference at each detector's output, we assume that there is no phase coherence between $V_{10}({\bf r},t), V_{20}({\bf r},t)$, so that $\langle V_{10}({\bf r},t)V_{20}({\bf r},t)\rangle$ $= 0, \langle V_{10}^*({\bf r},t)V_{20}({\bf r},t)\rangle = 0$.  In the case of fields of independent origins, this is automatically satisfied. On the other hand, the two fields may originate from one field $V_{0}({\bf r},t)$ via splitting by a beam splitter:  $V_{10}({\bf r},t) \propto V_{0}({\bf r},t), V_{20}({\bf r},t) \propto V_{0}({\bf r},t)e^{i\varphi}$, as shown in Fig.\ref{fig:sch}(c). In this case, we introduce a random phase $\varphi$ in field $V_{20}$ that averages out the second-order interference between $V_{10}$ and $V_{20}$.

For simplicity without loss of generality, we assume the fields are one-dimensional so we can absorb the position variable $z$ with time and only consider the temporal variable $t$. Then, the general format of the fourth-order quantities are in the form of $\langle V_{10}^*(t_1)V_{20}^*(t_2) V_{10}(t_1')V_{20}(t_2')\rangle$. In particular, cross terms like $\langle V_{10}^*(t_{1})V_{20}^*(t_{2}) V_{10}(t_{2})V_{20}(t_{1})\rangle$ and $\langle V_{10}^*(t_{1})V_{20}^*(t_{1}) V_{10}(t_{2})V_{20}(t_{2})\rangle$ result in fourth-order interference.

In order to obtain the interference terms mentioned above,  we introduce various delays $T_1,T_2,T_1',T_2'$ to account for different times of $t_1, t_2, t_1', t_2'$. Different values of $T_1,T_2,T_1',T_2'$ lead to different scenarios of interference. For example, when $T_1=T_1',T_2=T_2'$, this scheme is simply a Hong-Ou-Mandel interferometer for two fields $V_{10}, V_{20}$. When $V_{10}, V_{20}$ originate from $V_{0}$ as shown in Fig.\ref{fig:sch}(c), this scheme was shown to be able to measure the coherence time of $V_{0}$ \cite{ou88}. 

For the general scheme in Fig.\ref{fig:sch}(b), the fields at two detectors can be expressed as
\begin{eqnarray}\label{BS}
V(t) &=& \big [V_{10}(t+T_1) + V_{20}(t+T_2)\big]/\sqrt{2}, \cr  V'(t) &=& \big[ V_{10}(t+T_1') -  V_{20}(t+T_2')\big]/\sqrt{2}.
\end{eqnarray}
The coincidence measurement is related to $\langle I(t)I'(t+\tau)\rangle$ with
\begin{eqnarray}\label{I}
I &\equiv & |V(t)|^2 = |V_{10}|^2 + |V_{20}|^2 + V_{10}^*V_{20} + V_{20}^*V_{10},\cr
I' &\equiv & |V'(t+\tau)|^2 = |V_{10}'|^2 + |V_{20}'|^2 -{V_{10}^* }'V_{20}' - {V_{20}^*}'V_{10}'~~~~~~
\end{eqnarray}
where $V_{10}\equiv V_{10}(t+T_1), V_{20}\equiv V_{20}(t+T_2), V_{10}'\equiv V_{10}(t+T_1'+\tau), V_{20}'\equiv V_{20}(t+T_2'+\tau)$. Expanding $\langle I(t)I'(t+\tau)\rangle$ and keeping in mind the random phase $e^{i\varphi}$, we have
\begin{eqnarray}\label{II}
&&\langle I(t)I'(t+\tau)\rangle \cr
&&\hskip 0.2in = \langle (|V_{10}|^2 + |V_{20}|^2)(|V_{10}'|^2 + |V_{20}'|^2)\rangle \cr
&&\hskip 0.5in - \langle (V_{10}^*V_{20} + V_{20}^*V_{10})({V_{10}^* }'V_{20}' + {V_{20}^*}'V_{10}')\rangle,~~~~
\end{eqnarray}
where because of the random phase $e^{i\varphi}$, the un-paired cross terms like $\langle |V_{10}|^2 V_{20}^*V_{10}\rangle$ etc are zero. Expanding Eq.(\ref{II}), we have
\begin{eqnarray}\label{II-2}
&&\langle I(t)I'(t+\tau)\rangle \cr
&&\hskip 0.3in = \langle I_{10}I_{10}'\rangle + \langle I_{20}I_{20}'\rangle + \langle I_{10}I_{20}'\rangle + \langle I_{20}I_{10}'\rangle \cr
&&\hskip 0.6in
- (\langle V_{10}^*V_{20} {V_{20}^* }'V_{10}' \rangle +c.c.)\cr
&&\hskip 0.6in
- (\langle V_{10}^*V_{20} {V_{10}^* }'V_{20}' \rangle +c.c.),~~~~
\end{eqnarray}
where $c.c.$ means complex conjugate. Obviously, $\langle V_{10}^*$ $V_{20}{V_{20}^* }'V_{10}' \rangle$ and $\langle V_{10}^*V_{20} {V_{10}^* }'V_{20}' \rangle$ are the interference terms. Again, because of the random phase $e^{i\varphi}$, term $\langle V_{10}^*V_{20} {V_{10}^* }'V_{20}' \rangle$ and its complex conjugate are normally zero. The non-zero term can be explicitly written as
\begin{eqnarray}\label{V-cross}
&&\langle V_{10}^*V_{20} {V_{20}^* }'V_{10}' \rangle \cr
&&\hskip 0.4in = \langle V_{10}^*(t+T_1)V_{20}(t+T_2) \cr
&&\hskip 0.8in \times V_{20}^*(t+T_2'+\tau) V_{10}(t+T_1'+\tau) \rangle.
\end{eqnarray}

The evaluation of the non-vanishing interference term in Eq.(\ref{V-cross}) requires the knowledge of the statistics of field fluctuations. For example, Gaussian statistics of thermal fields will break the four-term average into two-term average: $\langle V_{10}^*V_{20} {V_{20}^* }'V_{10}' \rangle_{th} = \langle V_{10}^*V_{20} \rangle \langle{V_{20}^* }'V_{10}' \rangle + \langle V_{10}^*V_{10}' \rangle \langle V_{20} {V_{20}^* }' \rangle + \langle V_{10}^* {V_{20}^* }' \rangle \langle V_{20}  V_{10}' \rangle$. But we cannot go further for general fields without some approximations. Next, we will consider those approximations that lead to different scenarios in fourth-order interference.

\section{Various Scenarios}

\subsection{Scenarios with different field correlations}

The easiest approximation is to assume that there is no correlation between $V_{10}$ and $ V_{20}$ fields.  Then, Eq.(\ref{V-cross}) becomes
\begin{eqnarray}\label{V-cross1}
&&\langle V_{10}^*V_{20} {V_{20}^* }'V_{10}' \rangle \cr
&&\hskip 0.4in = \langle V_{10}^*(t+T_1)V_{10}(t+T_1'+\tau) \rangle \cr
&&\hskip 0.8in \times \langle V_{20}(t+T_2) V_{20}^*(t+T_2'+\tau)\rangle \cr  
&&\hskip 0.4in = I_{10}I_{20}\gamma_{11}(\tau - \Delta T_1) \gamma_{22}^*(\tau - \Delta T_2)
\end{eqnarray}
with $I_{j0}\equiv \langle |V_{j0}(t)|^2\rangle$, $\Delta T_j \equiv T_j-T_j' $, and 
\begin{eqnarray}\label{gma}
\gamma_{jj}(\tau) \equiv \langle V_{j0}^*(t)V_{j0}(t+\tau) \rangle/I_{j0}, 
\end{eqnarray}
where $j=1,2$. With this, Eq.(\ref{II-2}) becomes
\begin{eqnarray}\label{II-22}
&&\langle I(t)I'(t+\tau)\rangle \cr
&&\hskip 0.3in = I_{10}^2(1+\lambda_1) + I_{20}^2(1+\lambda_2)  \cr
&&\hskip 0.6in
+ 2 I_{10}I_{20} [1 - |\gamma_{11}\gamma_{22}|\cos(\varphi_{11}-\varphi_{22})],~~~~
\end{eqnarray}
where $\lambda_j\equiv \langle I_jI_j'\rangle/I_{j0}^2-1~(j=1,2)$ is the normalized auto-intensity correlation, describing the intensity fluctuations. $|\gamma_{jj}|, \varphi_{jj}$ are the magnitude and phase of $\gamma_{jj}$.

This scenario occurs when $V_{10}$ and $ V_{20}$ come from independent sources such as two celestial objects in the sky. $\varphi_{11}-\varphi_{22}$ contains information to resolve these two objects as in two-photon amplitude astronomy \cite{astro1}.

Another scenario is when $ V_{20}$ is from an ultra-stable coherent source or coherent state, that is $ V_{20} = \alpha$. In this case, $ V_{20}$ can be thought of as a weak local oscillator and it was first discussed in the context of single-photon nonlocality \cite{tan}. An application is in optical stellar interferometry for astronomy \cite{astro2}. We treat the case in the following.

\begin{figure}[t]
	\includegraphics[width=6cm]{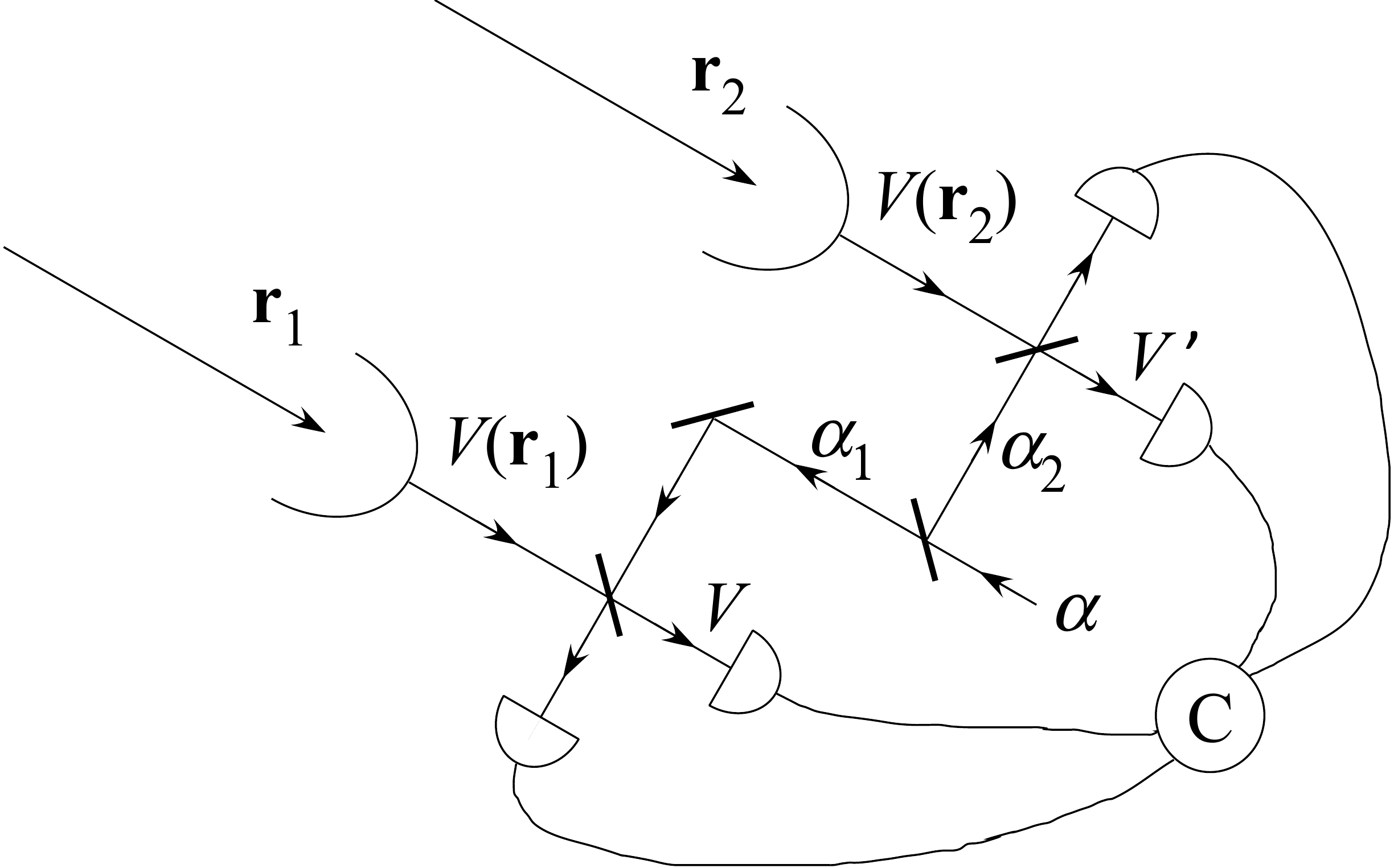}
	\caption{Schemes of fourth-order interference for application in astronomy }
	\label{fig:astro}
\end{figure}

The goal of stellar interferometry is to measure the normalized second-order coherence function \cite{bw} $\gamma({\bf r}_1,{\bf r}_2,\tau) \equiv \langle V^*({\bf  r}_1, t) V({\bf  r}_2, t+\tau)\rangle/\sqrt{\langle |V({\bf  r}_1, t)|^2 \rangle\langle |V({\bf  r}_2, t)|^2 \rangle}$  of the stellar optical field $V({\bf r}, t)$ at two locations ${\bf r}_1,{\bf r}_2$. Knowledge of  $\gamma({\bf r}_1,{\bf r}_2,\tau) $ for a large separation of ${\bf r}_1,{\bf r}_2$ will lead to high angular resolution by a Fourier transformation \cite{astro2}. Denote the incoming field at the two locations as $V(t)\equiv V({\bf  r}_1, t), \bar V(t)\equiv V({\bf  r}_2, t)$, which are equivalent to $V_1(t+T_1), V_1(t+T_1')$ with different delays in Fig.\ref{fig:sch}(b). We mix them with  local oscillator fields denoted by $\alpha_1,\alpha_2$, respectively, which are split from a common stable source of $\alpha$ (equivalent to $V_{20}$ in Fig.\ref{fig:sch}), as shown in Fig.\ref{fig:astro}. So the fields at the detectors are
\begin{eqnarray}\label{As}
V(t) = V(t) + \alpha_1,~~V'(t) = \bar V(t) + \alpha_2.
\end{eqnarray}
The delay of wave front between the two detectors is included in positions ${\bf r}_1,{\bf r}_2$. Intensity correlation measurement gives
\begin{eqnarray}\label{As2}
&& \langle I(t) I'(t+\tau)\rangle = \langle |V(t)|^2| V'(t+\tau) |^2\rangle\cr
&&\hskip 0.3in = \langle |V(t)|^2| \bar V(t+\tau) |^2\rangle  +I|\alpha_2|^2 + \bar I|\alpha_1|^2 \cr &&\hskip 0.8in   +|\alpha_1\alpha_2|^2 + \langle V^*(t) \bar V(t+\tau) \rangle \alpha_1\alpha_2^* \cr &&\hskip 1.4 in  + \langle V(t)\bar V^*(t+\tau) \rangle \alpha_1^*\alpha_2\cr &&\hskip 0.3in  =
I \bar I [1+\bar \lambda(\tau)] + |\alpha_1\alpha_2|^2  \cr &&\hskip 0.6in   +(I|\alpha_2|^2 + \bar I|\alpha_1|^2)\cr &&\hskip 0.8 in \times [1+ \xi|\gamma(\tau)|\cos(\varphi_{\gamma}+\Delta\phi_{\alpha})],~~~~
\end{eqnarray}
where $\xi \equiv 2|\alpha_1\alpha_2|\sqrt{I\bar I}/(I|\alpha_2|^2 + \bar I|\alpha_1|^2)$ with $I\equiv \langle |V|^2| \rangle, \bar I\equiv \langle |\bar V|^2| \rangle$, $\gamma(\tau) \equiv \gamma({\bf r}_1,{\bf r}_2, \tau)$, $e^{i\varphi_{\gamma}}\equiv \gamma/|\gamma|$,  $\Delta\phi_{\alpha} \equiv \phi_{\alpha_2}-\phi_{\alpha_1}$, and $1+ \bar \lambda(\tau) \equiv  \langle |V(t)|^2| \bar V(t+\tau) |^2\rangle/ I \bar I$. In deriving Eq.(\ref{As2}), we assume $\alpha_{1,2}$ has stable phases and the incoming fields $V, \bar V$ has random phases. Normally, stellar fields have $I=\bar I$ and are of thermal nature, so $\bar \lambda(\tau) = |\gamma(\tau)|^2$. Setting $|\alpha_1|^2=|\alpha_2|^2=I=\bar I$ in Eq.(\ref{As2}), we have
\begin{eqnarray}\label{As3}
&& \langle I(t) I'(t+\tau)\rangle = I^2(4+|\gamma|^2)[1+ {\cal V}\cos(\varphi_{\gamma}+\Delta\phi_{\alpha})],~~~~~~~
\end{eqnarray}
where ${\cal V} \equiv 2|\gamma(\tau)|/(4+|\gamma(\tau)|^2)$.

With stable local oscillators $\alpha_1, \alpha_2$, we can measure complex quantity $\gamma(\tau)$ from the two-photon interference fringe to achieve stellar interferometry in astronomy. Note that this scheme is similar to homodyne measurement technique in stellar interferometry but  photon counting technique is used here to avoid the shot noise problem \cite{astro3}. However, this method requires time resolution of the detectors better than coherence time in order to measure $\gamma(\tau)$ and thus limits the bandwidth, in a similar way to intensity interferometry \cite{HBT}.

\subsection{Scenarios with different delays}

 All random variables have some correlation time beyond which the fields are not related anymore. So, depending on the relationship between $T_1,T_2,T_1',T_2'$ as compared to coherence time $T_c$ of the fields and resolving time $T_R$ of detectors, we can make some approximations and have different scenarios of fourth-order interference, which give rise to different applications. We categorize them as follows:
\vskip 0.1in
\noindent (i) $T_1\sim T_2$ and $T_1'\sim T_2'$, but $|(T_1,T_2)-(T_1',T_2')|\gg T_c, T_R$. This is exactly the scenario depicted in Fig.\ref{fig:sch}(b). In this case, quantities $V_{10}^*(t+T_1)V_{20}(t+T_2)$ and $V_{20}^*(t+T_2'+\tau) V_{10}(t+T_1'+\tau)$ are well separated in time beyond any correlation time of the fields so that they are independent and we have
\begin{eqnarray}\label{uncorr}
&&\langle V_{10}^*(t+T_1)V_{20}(t+T_2) V_{20}^*(t+T_2'+\tau) V_{10}(t+T_1'+\tau) \rangle\cr
&&\hskip 0.4in \approx \langle V_{10}^*(t+T_1)V_{20}(t+T_2) \rangle \cr
&&\hskip 0.9in \times \langle V_{20}^*(t+T_2'+\tau) V_{10}(t+T_1'+\tau) \rangle \cr
&&\hskip 0.4in = \Gamma_{12}(\Delta T) \Gamma_{12}^*(\Delta T'),
\end{eqnarray}
where $\Delta T\equiv T_2-T_1, \Delta T'\equiv T_2'-T_1'$ and $\Gamma_{12}(\Delta T)\equiv \langle V_{10}^*(t+T_1)V_{20}(t+T_2) \rangle$. Notice that this term is normally zero because we assume that there is no coherence between $V_{10}$ and $V_{20}$ or we introduce a random phase between them in the case of common origin so that $\Gamma_{12}(\Delta T)=0$. But in the latter case, the random phase is canceled in the product of $\Gamma_{12}(\Delta T) \Gamma_{12}^*(\Delta T')$, as long as the phase changes slowly within the time period of $|T_1-T_1'|$. So, we will keep this term for the case when $V_{10}$ and $V_{20}$ are from a common origin as shown in Fig.\ref{fig:sch}(c). Moreover, because $|(T_1,T_2)-(T_1',T_2')|\gg T_c, T_R$, there is no intensity correlation between un-primed quantities and primed quantities, that is, $\langle I_{i0}I_{j0}'\rangle \approx \langle I_{i0}\rangle\langle I_{j0}'\rangle =   I_{i0}I_{j0}$. So, with the definition of $\gamma_{12} \equiv \Gamma_{12}/I_{10}I_{20}$, the overall coincidence measurement result is
\begin{eqnarray}\label{II-1}
&&\langle I(t)I'(t+\tau)\rangle \cr
&&\hskip 0.1in = I_{10}^2 + I_{20}^2 + 2 I_{10}I_{20}  - I_{10}I_{20}[\gamma_{12}(\Delta T) \gamma_{12}^*(\Delta T')+c.c.]\cr
&&\hskip 0.1in =  I_{10}^2 + I_{20}^2 + 2 I_{10}I_{20} \Big[1 - |\gamma_{12}(\Delta T) \gamma_{12}^*(\Delta T')| \cr &&\hskip 1.1 in \times\cos \big(\omega(\Delta T-\Delta T')+\Delta \varphi\big)\Big].
\end{eqnarray}
This gives rise to fourth-order interference.
Notice that the interference fringe does not depend on $\tau$.

\begin{figure}[t]
	\includegraphics[width=8cm]{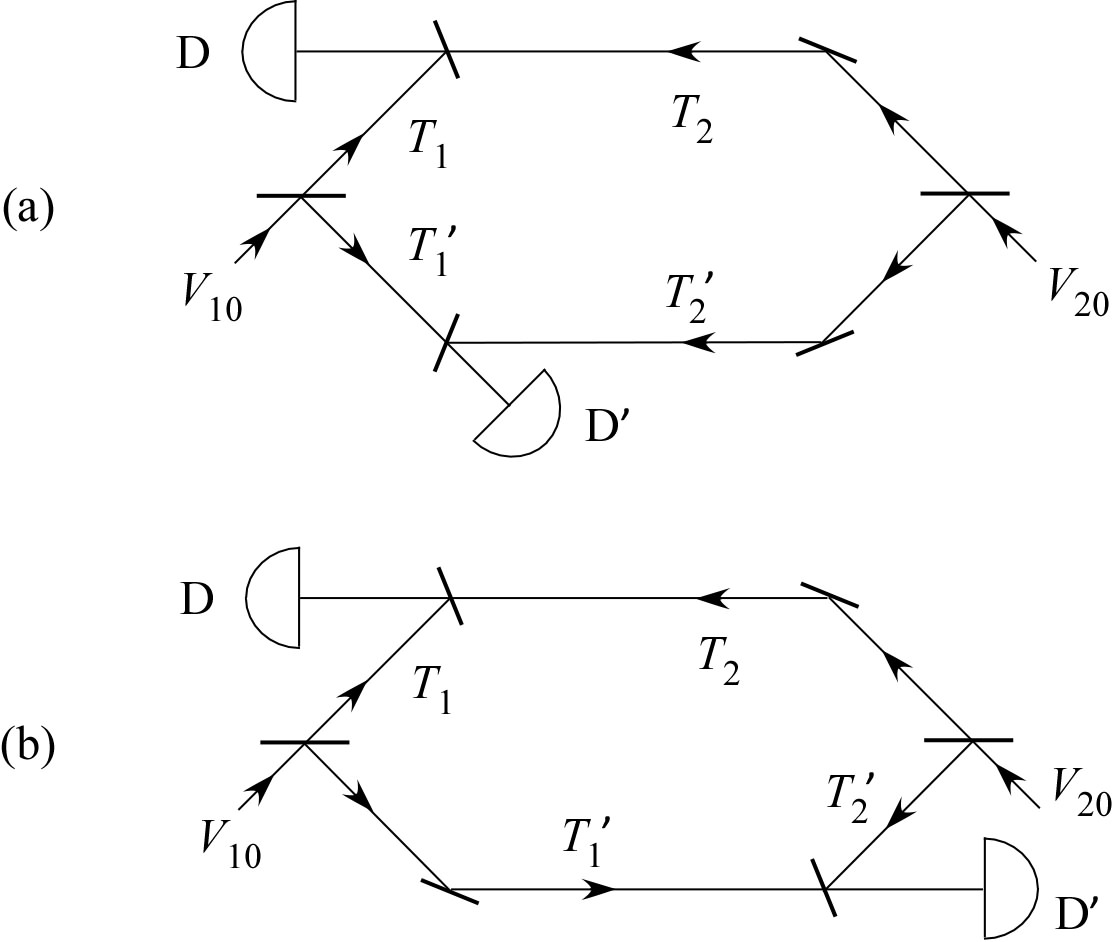}
	\caption{Schemes of fourth-order interference between two fields with (a) $T_1\sim T_1'$ and $T_2\sim T_2'$, but $|(T_1,T_1')-(T_2,T_2')|\gg T_c, T_R$ (Scenario (ii)); (b) $T_1\sim T_2'$ and $T_2\sim T_1'$, but $|(T_1,T_2')-(T_2,T_1')|\gg T_c, T_R$ (Scenario (iii)). }
	\label{fig:sch2}
\end{figure}

\vskip 0.1in
\noindent (ii) $T_1\sim T_1'$ and $T_2\sim T_2'$, but $|(T_1,T_1')-(T_2,T_2')|\gg T_c, T_R$. This scenario is depicted in Fig.\ref{fig:sch2}(a). Similar to scenario (i), we have
\begin{eqnarray}\label{uncorr2}
&&\langle V_{10}^*(t+T_1)V_{20}(t+T_2) V_{20}^*(t+T_2'+\tau) V_{10}(t+T_1'+\tau) \rangle\cr
&&\hskip 0.4in \approx \langle V_{10}^*(t+T_1)V_{10}(t+T_1'+\tau) \rangle \cr
&&\hskip 0.9in \times \langle V_{20}^*(t+T_2'+\tau) V_{20}(t+T_2) \rangle \cr
&&\hskip 0.4in = \Gamma_{11}(\Delta T_1+\tau) \Gamma_{22}^*(\Delta T_2+\tau),
\end{eqnarray}
where $\Delta T_1\equiv T_1'-T_1, \Delta T_2\equiv T_2'-T_2$ and $\Gamma_{jj}(\Delta T_j)\equiv \langle V_{j0}^*(t+T_j)V_{j0}(t+T_j') \rangle$.

There is no need for random phase $e^{i\varphi}$ in this scenario since the second-order coherence $\Gamma_{12}=0$ for $|(T_1,T_1')-(T_2,T_2')|\gg T_c, T_R$. So, the overall coincidence measurement result is
\begin{eqnarray}\label{II-3}
&&\langle I(t)I'(t+\tau)\rangle \cr
&&\hskip 0.2in = I_{10}^2(1+\lambda_1) + I_{20}^2(1+\lambda_2) +  2 I_{10}I_{20} \cr
&&\hskip 0.6in - I_{10}I_{20}[\gamma_{11}(\Delta T_1+\tau) \gamma_{22}^*(\Delta T_2+\tau)+c.c.]\cr
&&\hskip 0.2in = I_{10}^2(1+\lambda_1) + I_{20}^2(1+\lambda_2)  \cr
&&\hskip 0.6in + 2 I_{10}I_{20}\big [1 - |\gamma_{11}(\Delta T_1+\tau) \gamma_{22}(\Delta T_2+\tau)|\cr
&&\hskip 1 in \times \cos (\omega (\Delta T_1-\Delta T_2)+\Delta\varphi)\big],
\end{eqnarray}
where $\lambda_j \equiv \langle I_{j0}I_{j0}'\rangle/I_{j0}^2 -1 (j=1,2)$ describes the intensity fluctuation of each field. This again shows fourth-order interference.
A special case is when $T_1'=T_1, T_2'=T_2$ and two BS's before the detectors merge into one. This is an un-balanced Mach Zehnder interferometer (MZI) if the two fields are from the splitting of one field (Fig.\ref{fig:sch}(c)) or a classical version of the HOM interferometer if the two fields are independent. In this case, we have
\begin{eqnarray}\label{II-3a}
&&\langle I(t)I'(t+\tau)\rangle \cr
&&\hskip 0.5in = I_{10}^2(1+\lambda_1) + I_{20}^2(1+\lambda_2) +  2 I_{10}I_{20} \cr
&&\hskip 1in - I_{10}I_{20}[\gamma_{11}(\tau) \gamma_{22}^*(\tau)+c.c.],
\end{eqnarray}
where the interference is in the form of a dip as the delay $\Delta T\equiv |T_1-T_2|$ or detector time delay $\tau$ is scanned. Note that for thermal field of the same kind with, we have $\lambda_1=\lambda_2=|\gamma_{11}(\tau)|^2=|\gamma_{22}|^2$ and  $I_{10}=I_{20}\equiv I_0$. Then Eq.(\ref{II-3a}) becomes
\begin{eqnarray}\label{II-3b}
\langle I(t)I'(t+\tau)\rangle   = 4I_{0}^2,
\end{eqnarray}
showing no interference because the bunching effect of the thermal fields cancels the HOM destructive interference effect. This is only true for stationary thermal fields. In the case of non-stationary pulsed thermal fields, the situation is different because of the requirement of pulse overlap (see later).

\vskip 0.1in
\noindent (iii) $T_1\sim T_2'$ and $T_2\sim T_1'$, but $|(T_1,T_2')-(T_2,T_1')|\gg T_c, T_R$. This scenario is depicted in Fig.\ref{fig:sch2}(b) and we have
\begin{eqnarray}\label{uncorr3}
&&\langle V_{10}^*(t+T_1)V_{20}(t+T_2) V_{20}(t+T_2'+\tau) V_{10}^*(t+T_1'+\tau) \rangle\cr
&&\hskip 0.4in \approx \langle V_{10}^*(t+T_1)V_{20}(t+T_2'+\tau) \rangle \cr
&&\hskip 0.9in \times \langle V_{20}(t+T_2) V_{10}^*(t+T_1'+\tau) \rangle \cr
&&\hskip 0.4in = \Gamma_{12}(\Delta \bar T_1'+\tau) \Gamma_{21}^*(\tau-\Delta \bar T_2'),
\end{eqnarray}
where $\Delta \bar T_1' \equiv T_2'-T_1, \Delta \bar T_2' \equiv T_2-T_1'$. Similar to scenario (i), this is for the case of two fields with a common origin. But
in this case, we cannot have random phase $e^{i\varphi}$ because otherwise, the term above will be zero. On the other hand, since $|(T_1,T_2')-(T_2,T_1')|\gg T_c, T_R$, there is no second-order interference in D1 and D2 in any case so there is no need for the random phase. So, the result of coincidence measurement is
\begin{eqnarray}\label{II-4}
&&\langle I(t)I'(t+\tau)\rangle \cr
&&\hskip 0.2in = I_{10}^2 + I_{20}^2 +  2 I_{10}I_{20} \cr
&&\hskip 0.6in - I_{10}I_{20}[\gamma_{12}(\Delta \bar T_1'+\tau) \gamma_{21}^*(\tau-\Delta \bar T_2')+c.c.]\cr
&&\hskip 0.2in = I_{10}^2 + I_{20}^2  \cr
&&\hskip 0.6in + 2 I_{10}I_{20}\big [1 - |\gamma_{12}(\Delta \bar T_1'+\tau) \gamma_{21}^*(\tau-\Delta \bar T_2')|\cr
&&\hskip 1 in \times \cos (\omega (\Delta \bar T_1' +\Delta \bar T_2')+\Delta\varphi)\big].
\end{eqnarray}
This scenario is similar to the case of a classical Franson interferometer for thermal fields \cite{ihn17}.

Both the results in scenario (ii) and (iii) depend on $\tau$. So, a time-resolved coincidence measurement is required, which means $T_R\ll T_c$. But interference in these two scenarios will disappear if $T_R\gg T_c$, or detector's response is too slow to resolve the details of the field fluctuations. This was pointed out in Ref.\citenum{ou-mandel90} for classical Franson interferometer. However, this condition leads to  the following scenario.
\vskip 0.1in
\noindent (iv) $T_R\gg T_c$. Because of the slowness of the detectors, the result of coincidence is an average over detectors' resolving time $T_R$: $R_c = (1/T_R)\int_{T_R} d\tau \langle I(t)I'(t+\tau)\rangle $. This scenario was discussed in Ref.\citenum{ou88} where it was argued that all higher order correlations are averaged out due to slow detectors and the result of coincidence measurement is exactly the same as Eq.(\ref{II-1}), that is,
\begin{eqnarray}\label{II-42}
&&R_c =\frac{1}{T_R}\int_{T_R} d\tau \langle I(t)I'(t+\tau)\rangle \cr
&&\hskip 0.1in = I_{10}^2 + I_{20}^2 + 2 I_{10}I_{20}  - I_{10}I_{20}[\gamma_{12}(\Delta T) \gamma_{12}^*(\Delta T')+c.c.].\cr &&
\end{eqnarray}
Note that this scenario here does not assume anything for $T_1, T_2, T_1', T_2'$. But Eq.(\ref{II-4}) requires $|\Delta T|, |\Delta T'| \ll T_c$ in order to have non-zero interference terms. A
 special case is when $T_1=T_1', T_2=T_2'$ or $|\Delta T|= |\Delta T'|$. Under this condition, Eq.(\ref{II-1}) becomes
\begin{eqnarray}\label{II-5}
&&R_c  =  I_{10}^2 + I_{20}^2 + 2 I_{10}I_{20} \big[1 - |\gamma_{12}(\Delta T)|^2\big].
\end{eqnarray}
The fourth-order interference is in the form of a dip as the delay $\Delta T$ is scanned. This case is exactly the Mach-Zehnder interferometer scheme presented in Ref.\citenum{ou88}, which can be used to measure $|\gamma_{12}|$ and the coherence time of an incoming field independent of the photon statistics of the incoming field.

\section{The case of fields in pulse trains}

For a nonstationary field $V_1(t)$ in the form of
a quasi-continuous wave (quasi-cw) train of pulses, the situation somehow becomes relatively simple because the single pulse is usually much faster than the response of the detectors so that the result is a time integral of the single pulse profile. For this case, the field
can be written in general as
\begin{eqnarray}\label{E-pulse}
V_1(t)   =  \sum_j A_j f(t-j\Delta t),
\end{eqnarray}
where $f(t)$ is the normalized single
pulse profile with a pulse width $\delta t$, which we assume is the same for all pulses in the train, $A_j$ is the amplitude of the j-th pulse, and $\Delta t (\gg \delta t)$ is the
interval between two adjacent pulses. Here, we consider only one polarization and can treat the field as a scalar field. Then the instantaneous
intensity is
\begin{eqnarray}\label{I-pulse}
I_1(t) = |V_1(t)|^2   & =&  \sum_{j,k} A_j^*f^*(t-j\Delta t)A_k f(t-k\Delta t)\cr
& =&  \sum_{j} |A_j f(t-j\Delta t)|^2,
\end{eqnarray}
where the cross terms are zero because pulse width $\delta t$ is much smaller than the pulse separation $\Delta t$. The photo-current from the detector illuminated by this field is then
\begin{eqnarray}\label{i-pulse}
i_1(t) & =& \int dt' k(t-t') I_1(t') \cr
& =&  \sum_{j} |A_j|^2 \int dt' k(t-t')| f(t'-j\Delta t)|^2\cr
& =&  \sum_{j} |A_j|^2  k(t-j\Delta t),
\end{eqnarray}
where $k(t)$ is the detector's response function and we assume that single pulse width of $f(t)$ is much narrower than the detector's response function $k(t)$ so that we can pull $k(t)$ out of the integral. The average photo-current over a long time of $T (\gg \Delta t)$ is then
\begin{eqnarray}\label{i-average}
\langle i_1\rangle &=& \frac{1}{T}\int_T dt i_1(t) =  Q R_p \frac{1}{N} \sum_{j=1}^N |A_j|^2 \cr &=& Q R_p \langle |A_j|^2\rangle_j,~~~~
\end{eqnarray}
where $Q\equiv \int dt k(t)$ is the total charge produced in the detector for one pulse, $R_p$ is the pulse repetition rate, and $N=[T/\Delta t] = R_p T$ is the number of pulses in time $T$. $\langle \rangle_j$ is the average over the $N$ pulses. For later calculation, we need to evaluate auto-correlation of the photo-current within a coincidence window of $T_R$. The time average is given by
\begin{eqnarray}\label{R-auto}
R_{11} &=& \frac{1}{T} \int_T dt \int_{T_R} d\tau i_1(t)i_1(t+\tau)\cr
 &=& \frac{1}{T}\int_T dt \int_{T_R} d\tau \sum_{i,j} |A_i|^2 k(t-i\Delta t)\cr &&\hskip 1.3 in \times |A_j|^2 k(t+\tau -j\Delta t)\cr
  &=& \frac{1}{T}\int_T dt \int_{T_R} d\tau \sum_{j} |A_j|^4 k(t-j\Delta t)\cr &&\hskip 1.3 in \times k(t+\tau -j\Delta t)\cr
 &=&  R_p Q^2 \langle |A_j|^4 \rangle_j,
\end{eqnarray}
where we assume the detectors can resolve different pulses so that $T_R< \Delta t$ and $k(t -i\Delta t)k(t+\tau -j\Delta t)=0$ if $i\neq j$.

Suppose there is a second field $V_2(t)$ in a pulse train with the same pulse separation $\Delta t$:
\begin{eqnarray}\label{E2-pulse}
V_2(t)   =  \sum_j B_j g(t-j\Delta t),
\end{eqnarray}
where the amplitude of each pulse is denoted as $B_j$ and the pulse profile is $g(t)$. The coincidence measurement between the two fields is described by coincidence rate:
\begin{eqnarray}\label{R2}
R_{12} &=& \frac{1}{T} \int_T dt \int_{T_R} d\tau i_1(t)i_2(t+\tau)\cr
 &=&  R_p Q^2 \langle |A_j|^2|B_{j}|^2\rangle_j,
\end{eqnarray}
whose derivation is similar to Eq.(\ref{R-auto}).

Now, let us inject the two fields into the unbalanced interferometers shown in Figs.\ref{fig:sch},\ref{fig:sch2}. We consider again the different scenarios of delays as in the stationary case. But we write the delays in terms of pulse separation $\Delta t$: $T_1=N_1\Delta t+d_1/c, T_2=N_2\Delta t+d_2/c, T_1'=N_1'\Delta t+d_1'/c, T_2'=N_2'\Delta t+d_2'/c$ with $d_1,d_2,d_1',d_2' (< c\Delta t)$ being the extra path delay between two adjacent pulses.

With random phase relation between $V_1$ and $V_2$, similar to Eq.(\ref{II-2}) in the stationary case, the coincidence measurement between the two outputs of the interferometer is related to eight terms corresponding to two auto-correlation terms $\langle I_{1}I_{1}'\rangle, \langle I_{2}I_{2}'\rangle$, two cross-correlation terms $\langle I_{1}I_{2}'\rangle,\langle I_{2}I_{1}'\rangle$,  and four interference terms. Our discussion of these terms needs to involve detection processes for the case of pulse trains. For the four intensity correlation terms, their contributions to coincidence measurement can be evaluated in a similar way to Eqs.(\ref{R-auto},\ref{R2}) and have the form of
\begin{eqnarray}\label{R22}
R_{11'} &=&  R_p Q^2 \langle |A_{j+N_1}|^2|A_{j+N_1'}|^2\rangle_j,\cr
R_{22'} &=&  R_p Q^2 \langle |B_{j+N_2}|^2|B_{j+N_2'}|^2\rangle_j,\cr
R_{12'} &=&  R_p Q^2 \langle |A_{j+N_1}|^2|B_{j+N_2'}|^2\rangle_j,\cr
R_{1'2} &=&  R_p Q^2 \langle |A_{j+N_1'}|^2|B_{j+N_2}|^2\rangle_j.
\end{eqnarray}

The contributions from the four interference terms are more complicated to evaluate. Using Eqs.(\ref{E-pulse}, \ref{E2-pulse}) for $V_1, V_2$, we have
\begin{eqnarray}\label{V12}
&&\int dt' k(t-t') V_{1}^*(t'+T_1)V_{2}(t'+T_2) \cr
&&\hskip 0.2in = \int dt' k(t-t') \sum_{i,j} A_i^*  f^*(t'+T_1 -i\Delta t) \cr
&&\hskip 1.5 in \times B_j g(t'+T_2 -j\Delta t)\cr
&&\hskip 0.2in  = \int dt' k(t-t') \sum_{i,j} A_i^*  f^*(t'+d_1/c -(i-N_1)\Delta t)\cr
&&\hskip 0.7in \times B_j  g(t'+d_2/c -(j-N_2)\Delta t) \cr
&&\hskip 0.2in = \beta(\Delta d/c)\sum_j A_{j+N_1}^*B_{j+N_2} k(t-j\Delta t)
\end{eqnarray}
and
\begin{eqnarray}\label{V21-tau}
&&\int dt' k(t+\tau-t') V_{2}^*(t'+T_2') V_{1}(t'+T_1')  \cr
&&\hskip 0.2in = \int dt' k(t+\tau-t') \sum_{i,j} B_i^*  g^*(t'+T_2' -i\Delta t) \cr
&&\hskip 1.5 in \times A_j f(t'+T_1' -j\Delta t)\cr
&&\hskip 0.2in  = \int dt' k(t+\tau-t') \sum_{i,j} A_i  f(t'+d_1'/c -(i-N_1')\Delta t)\cr
&&\hskip 0.7in \times B_j^* g^*(t'+d_2'/c -(j-N_2')\Delta t) \cr
&&\hskip 0.2in = \beta^*(\Delta d'/c)\sum_j B_{j+N_2'}^* A_{j+N_1'} k(t+\tau-j\Delta t)
\end{eqnarray}
as parts of the contributions in the two detectors from the interference terms. Here, $\beta(\Delta d/c)\equiv \int dt f^*(t)g(t+\Delta d/c)$ with $\Delta d \equiv d_2-d_1$ and $\Delta d' \equiv d_2'-d_1'$.

The time average of the contribution of the first two interference terms to the overall coincidence is then
\begin{eqnarray}\label{R-1221}
R_{1221} &=& R_pQ^2 \beta(\Delta d/c) \beta^*(\Delta d'/c)\cr
 &&\hskip 0.1in \times \langle A_{j+N_1}^*B_{j+N_2}B_{j+N_2'}^* A_{j+N_1'}\rangle_j +c.c.~~~~
\end{eqnarray}
Similarly, the contribution of the last two interference terms is
\begin{eqnarray}\label{R-1212}
R_{1212} &=& R_pQ^2 \beta(\Delta d/c) \beta(\Delta d'/c)\cr
 &&\hskip 0.1in \times \langle A_{j+N_1}^*B_{j+N_2}A_{j+N_1'}^*B_{j+N_2'} \rangle_j +c.c.~~~~
\end{eqnarray}

From Eqs.(\ref{R2},\ref{R-1221},\ref{R-1212}), we sum up all the contributions to obtain the overall coincidence rate for the two outputs of the interferometer:
\begin{eqnarray}\label{R-1221-2}
R_{c} &=& R_pQ^2\Big\{\langle |A_{j+N_1}|^2|A_{j+N_1'}|^2\rangle_j+ \langle |B_{j+N_2}|^2|B_{j+N_2'}|^2\rangle_j\cr
&&\hskip 0.3in + \langle |A_{j+N_1}|^2|B_{j+N_2'}|^2\rangle_j+
 \langle |A_{j+N_1'}|^2|B_{j+N_2}|^2\rangle_j\cr
 &&\hskip 0.3 in -\big [\beta(\Delta d/c) \beta^*(\Delta d'/c) \cr
 &&\hskip 0.6in \times \langle A_{j+N_1}^*B_{j+N_2}B_{j+N_2'}^* A_{j+N_1'}\rangle_j + c.c.\big ]\cr
&&\hskip 0.3 in -\big [\beta(\Delta d/c) \beta(\Delta d'/c)\cr
  &&\hskip 0.6in \times \langle A_{j+N_1}^*B_{j+N_2}A_{j+N_1'}^*B_{j+N_2'} \rangle_j + c.c.\big ]\Big \}.
\cr &&
\end{eqnarray}
Compared to the stationary case in Eq.(\ref{II-2}), we find extra factors of $\beta(\Delta d/c),\beta(\Delta d'/c)$ in the interference terms. Since by Cauchy's inequality we have $|\beta(\tau)|^2 = |\int dt f^*(t)g(t+\tau)|^2 \le \int dt |f(t)|^2 \int dt |g(t+\tau)|^2= 1$, this factor requires the overlap of the pulses from the two input fields at each detector and thus arises from mode match of the temporal profiles of the two fields. It gives rise to the degree of indistinguishability for interference.

Besides the two mode matching factors, the pulsed case is the same as the stationary case and gives rise to the same three scenarios (i-iii). But we do not have the scenario (iv) since we already assume $\Delta t > T_R$.

The dependence of interference terms on the mode match factor can be used in remote sensing to probe the change of the temporal profile when one of the fields passes through a medium, which can cause the change in $f(t)$ or $g(t)$ and thus $\beta(\tau)$, which is related to the visibility of interference. In fact, this was recently demonstrated with an unbalanced Mach-Zehnder interferometer to characterize the influence of dispersion on the temporal modes of the pulses \cite{Zhao21}. This corresponds to scenario (ii) with $T_1=T_1'$, $T_2=T_2'$ or $\Delta T=\Delta T' \gg T_c, T_R$.

\section{Summary and discussion}

We discussed in this paper various scenarios in fourth-order interference where path differences between interfering fields are much larger than their coherence length. We find phase-sensitive interference fringes may occur in a number of scenarios even though there exist large path differences beyond coherence length.  The unbalanced nature of these interference phenomena should find applications in remote optical sensing by interferometric technique.

Although fourth-order correlations are considered, the visibility of interference still depends on second order correlation functions.  Especially, some scenarios require time-resolved two-photon coincidence measurement within the coherence time (Scenario A and  B(ii) and B(iii)). This indicates that these phenomena are in essence originated from second-order coherence either between the two interfering fields or within each field itself. Since the coherence time gives the size of the coherent wave packet in the stationary case, the requirement of time resolution is equivalent to the temporal mode match factor of $\beta(\Delta d)$ in the pulsed case. In some sense, the size of the coherent wave packet in the stationary case is equivalent to the size of the temporal mode in the pulsed case.

On the other hand, fourth-order correlations do contribute to the results by adding to the baseline in the form of intensity fluctuations in some cases (quantity $\lambda$ in Eqs.(\ref{II-22}, \ref{II-3})). Their effect is to reduce the visibility of interference, as shown in Eq.(\ref{As3}).

\vskip 0.4in

\noindent \textbf{Acknowledgment}\\

Xiaoying Li is supported by National Natural Science Foundation of China (Grant Nos. 91836302 and 12074283).


\vskip 0.1in




\begin{thebibliography}{32}%
\newcommand{\enquote}[1]{``#1''}

	\bibitem{opsens} J\"org Haus, {\it Optical Sensors: Basics and Applications}, Wiley-VCH, 2010.

	\bibitem{bw} M. Born and E. Wolf, {\it Principle of Optics}, Pergamon Press, 1st ed., 1959; 6th ed., 1980.
	
	\bibitem{hom} C. K. Hong, Z. Y. Ou, and L. Mandel, \enquote{Measurement of subpicosecond time intervals between two photons by interference,} Phys. Rev. Lett. {\bf 59}, 2044 (1987).
	
	\bibitem{ou89} Z. Y. Ou, E. C. Gage, B. E. Magill, and L. Mandel, \enquote{Fourth-order interference technique for determining the coherence time of a light beam,} J. Opt. Soc. Am. B \textbf{6}, 100 (1989).

	\bibitem{xyli-th} Xiaoying Li, Lei Yang, Liang Cui, Zhe Yu Ou, and Daoyin Yu, \enquote{Observation of quantum interference between a single-photon state and a thermal state generated in optical fibers,} Opt. Express {\bf 16}, 12505 (2008).

	\bibitem{Ma-th} X. Ma, L. Cui, and X. Li, \enquote{Hong-Ou-Mandel interference between independent sources of heralded ultrafast single photons: influence of chirp,} J. Opt. Soc. Am. B{\bf 32}, 946 (2015).
	
	\bibitem{fran} J. D. Franson, \enquote{Bell inequality for position and time,}  Phys. Rev. Lett. {\bf 62} 2205 (1989).

	\bibitem{ou90} Z. Y. Ou, X. Y. Zou, L. J. Wang, and L. Mandel, \enquote{Observation of nonlocal interference in separated photon channels,} Phys. Rev. Lett. {\bf 65} 321 (1990).
	
	\bibitem{kwi} P. G. Kwiat, W. A. Vareka, C. K. Hong, H. Nathel, and R. Y. Chiao, \enquote{Correlated two-photon interference in a dual-beam Michelson interferometer,} Phys. Rev. A {\bf 41}, 2910 (1990).

	\bibitem{fran-ex91} J. Brendel, E. Mohler, and W. Martienssen, \enquote{Time-resolved dual-beam two-photon interferences with high visibility,} Phys. Rev. Lett. {\bf 66}, 1142 (1991).

	\bibitem{ou-mandel90} Z. Y. Ou and L. Mandel, \enquote{Classical Treatment of the Franson Two-Photon Correlation Experiment,} J. Opt. Soc. Am. B{\bf 7}, 2127 (1990).

	\bibitem{njp16} Vincenzo Tamma and Johannes Seiler, \enquote{Multipath correlation interference and controlled-NOT gate simulation with a thermal source,} New J. Phys. \textbf{18}, 032002 (2016).

	\bibitem{ihn17} Yong Sup Ihn, Yosep Kim, Vincenzo Tamma, and Yoon-Ho Kim, \enquote{Second-Order Temporal Interference with Thermal Light: Interference beyond the Coherence Time,} Phys. Rev. Lett. {\bf 119}, 263603 (2017).

	\bibitem{zhouqiang} Yun-Ru Fan, Chen-Zhi Yuan, Rui-Ming Zhang, Si Shen, Peng Wu, He-Qing Wang, Hao Li, Guang-Wei Deng, Hai-Zhi Song, Li-Xing You, Zhen Wang, You Wang, Guang-Can Guo, and Qiang Zhou, \enquote{Effect of dispersion on indistinguishability between single-photon wave-packets,} Photonics Res. \textbf{9}, 1134 (2021).

	\bibitem{astro1} Paul Stankus,  Andrei Nomerotski, and Anze  Slosar, \enquote{Two-photon amplitude interferometry for precision astrometry,} arXiv:2010.09100 (2020).

	\bibitem{tan} S. M. Tan, D. F. Walls, and M. J. Collett, \enquote{Nonlocality of
a single photon,} Phys. Rev. Lett. {\bf 66}, 252 (1991).

	\bibitem{astro2} John D. Monnier, \enquote{Optical interferometry in astronomy,} Rep. Prog. Phys. {\bf 66}, 789 (2003).

	\bibitem{astro3} M. A. Johnson, A. L. Betz, and C. H. Townes,  \enquote{10-micron heterodyne stellar interferometer,} Phys. Rev. Lett. {\bf 33} 1617 (1974).

	\bibitem{HBT} R. Hanbury-Brown and R. Q. Twiss, \enquote{A test of a new type of
stellar interferometer on Sirius,} Nature {\bf 178}, 1046 (1956).

	\bibitem{ou88} Z. Y. Ou, E. C. Gage, B. E. Magill, and L. Mandel, \enquote{Fourth-order interference technique for determining the coherence time of a light beam,} J. Opt. Soc. Am. B \textbf{6}, 100 (1988).

	\bibitem{Zhao21} Wen Zhao, Nan Huo, Liang Cui, and Xiaoying Li, and Z. Y. Ou, \enquote{Propagation of temporal mode multiplexed optical fields in fibers: influence of dispersion,} arXiv (2021).

\end{thebibliography}
\end{document}